\documentclass[floatfix,nofootinbib,twocolumn,showpacs,superscriptaddress, reprintnumbers,amssymb,letterpaper,amsmath,pra]{revtex4-2}

\usepackage[scientific-notation=true]{siunitx}
\usepackage{graphicx}
\usepackage{dcolumn}
\usepackage{bm}
\usepackage{xcolor}
\usepackage[colorlinks=true,allcolors=blue]{hyperref}
\usepackage{amsmath}
\usepackage{lineno}
\usepackage{float}
\usepackage{tablefootnote}
\usepackage{scrextend}
\usepackage{braket}
\usepackage{calrsfs}
\usepackage[english]{babel}
\DeclareMathAlphabet{\pazocal}{OMS}{zplm}{m}{n}

\usepackage{tabstackengine}
\stackMath

\usepackage{array}
\newcolumntype{C}[1]{>{\centering\arraybackslash}p{#1}}

\begin{document}
\title
{
Deployment and validation of predictive 6-dimensional beam diagnostics through generative reconstruction with standard accelerator elements
}

\author{Seongyeol Kim}\email{sykim12@postech.ac.kr}
\affiliation{Pohang Accelerator Laboratory, POSTECH, Pohang, Gyeongsangbuk-do 37673, Republic of Korea}
\author{Juan Pablo Gonzalez-Aguilera}
\affiliation{Department of Physics and Enrico Fermi Institute, University of Chicago, Chicago, IL 60637, USA}
\author{Ryan Roussel}\email{rroussel@slac.stanford.edu}
\affiliation{SLAC National Accelerator Laboratory, Menlo Park, CA 94025, USA}
\author{Gyujin Kim}
\affiliation{Pohang Accelerator Laboratory, POSTECH, Pohang, Gyeongsangbuk-do 37673, Republic of Korea}
\author{Auralee Edelen}
\affiliation{SLAC National Accelerator Laboratory, Menlo Park, CA 94025, USA}
\author{Myung-Hoon Cho}
\affiliation{Pohang Accelerator Laboratory, POSTECH, Pohang, Gyeongsangbuk-do 37673, Republic of Korea}
\author{Young-Kee Kim}
\affiliation{Department of Physics and Enrico Fermi Institute, University of Chicago, Chicago, IL 60637, USA}
\author{Chi Hyun Shim}
\affiliation{Pohang Accelerator Laboratory, POSTECH, Pohang, Gyeongsangbuk-do 37673, Republic of Korea}
\author{Hoon Heo}
\affiliation{Pohang Accelerator Laboratory, POSTECH, Pohang, Gyeongsangbuk-do 37673, Republic of Korea}
\author{Haeryong Yang}\email{highlong@postech.ac.kr}
\affiliation{Pohang Accelerator Laboratory, POSTECH, Pohang, Gyeongsangbuk-do 37673, Republic of Korea}


\begin{abstract}

Understanding the 6-dimensional phase space distribution of particle beams is essential for optimizing accelerator performance.
Conventional diagnostics such as use of transverse deflecting cavities offer detailed characterization but require dedicated hardware and space. 
Generative phase space reconstruction (GPSR) methods have shown promise in beam diagnostics, yet prior implementations still rely on such components.
Here we present the first experimental implementation and validation of the GPSR methodology, realized by the use of standard accelerator elements including accelerating cavities and dipole magnets, to achieve complete 6-dimensional phase space reconstruction.
Through simulations and experiments at the Pohang Accelerator Laboratory X-ray Free Electron Laser facility, we successfully reconstruct complex, nonlinear beam structures.
Furthermore, we validate the methodology by predicting independent downstream measurements excluded from training, revealing near-unique reconstruction closely resembling ground truth. 
This advancement establishes a pathway for predictive diagnostics across beamline segments while reducing hardware requirements and expanding applicability to various accelerator facilities.

\end{abstract}

\maketitle
Particle accelerator-based user facilities are indispensable tools that power scientific discovery in the biological, chemical and material sciences \cite{Seddon_2017,HUANG2021100097}.
Understanding the precise distribution of particles in an accelerated beam is therefore critical to improving and maintaining optimal performance of these facilities.
For example, X-ray Free Electron Laser (XFEL) facilities are strongly dependent on the distribution of electrons in 6-dimensional position-momentum space to generate high-brightness, ultra-short X-ray pulses~\cite{PhysRevLett.85.3825,PhysRevLett.80.289,mhz_xfel_superconducting,kang2017hard,nam2021high}.
The precise structure of the beam distribution during the acceleration and compression stages of the accelerator impacts overall beam quality due to collective effects such as space charge and coherent synchrotron radiation (CSR)~\cite{SALDIN1997373,DIMITRI20141,PhysRevAccelBeams.19.100703,PhysRevAccelBeams.20.054402}, that in-turn, negatively impacts FEL performance. 
As a result, methods are needed to efficiently and precisely characterize 6-dimensional phase space distribution at multiple locations along the beamline.

While methods for measuring lower dimensional phase space distribution structure of particle beams are well understood, either through direct measurements including sub-sampling portions of the beam distribution~\cite{zhang1996emittance} or via indirect measurements such as tomographic manipulations~\cite{MCKEE1995264, PhysRevSTAB.6.122801, HOCK20138}, their applicability becomes limited on 5-~\cite{PhysRevAccelBeams.27.072801,jastermerz2025experimentaldemonstrationtomographic5d} or 6-dimensional~\cite{PhysRevLett.121.064804} phase space distributions due to largely increased number of samples required~\cite{hoover2022towards}. 
Previous attempts at characterizing the high-dimensional phase space distribution of beams require a substantial amount of dedicated experimental beam time to conduct ($>$24 hrs for 5- and 6-dimensional characterization~\cite{jastermerz2025experimentaldemonstrationtomographic5d, PhysRevLett.121.064804}), making these methods impractical for use at user facilities. 
Additionally, these methods also require the use of specialized diagnostic elements to characterize the beam distribution, such as movable slits or transverse deflecting cavities (TCAVs)~\cite{emma2000transverse,Marchetti2021}. 
These additional requirements can add complexity and operational costs to diagnose the beam at multiple locations and reduce high-value space in the accelerator beamline.

Recently, a new machine learning technique, known as generative phase space reconstruction (GPSR), has been developed to address the limitations of conventional beam diagnostic methods~\cite{PhysRevLett.130.145001}.
This approach is different from virtual diagnostics which have also been actively investigated~\cite{7454846,edelen2018opportunitiesmachinelearningparticle,PhysRevAccelBeams.21.112802,PhysRevAccelBeams.23.044601,PhysRevLett.121.044801,scheinker2021adaptive,scheinker2024cdvae}.
The main difference between the GPSR method and virtual diagnostics is that it reconstructs the input beam distribution  rather than predicting the output for a new set of input parameters.
For the case of the generative model for phase space diagnostics, studies on normalizing flows~\cite{PhysRevResearch.6.033163,zl2h-3v32} are also actively carried out in recent years.

\begin{figure*}[ht]
   \centering
   \includegraphics[width=17.5 cm]{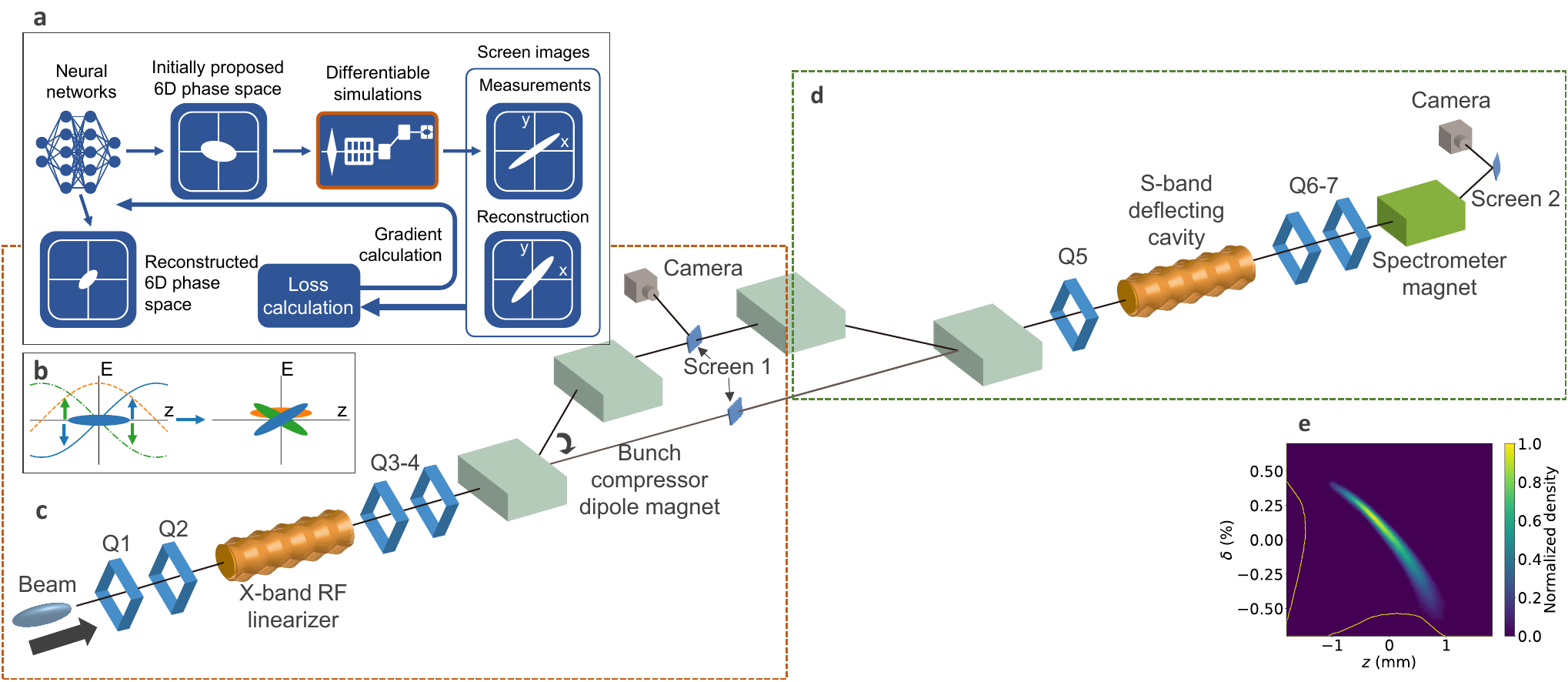}
   \caption{\textbf{a}. Schematics of the GPSR method~\cite{PhysRevLett.130.145001}. 
   \textbf{b}: Cartoon showing an evolution of beam energy chirp by the accelerating field.
            \textbf{c}. Schematic view of the first bunch compressor~\cite{diMitri2018beam} section at the PAL-XFEL.
            Training and test datasets are measured at the screen~1~\cite{walasek2011scintillating, Kim:IPAC2016-WEYB01}.
            The lattice used for the reconstruction training process is in the dashed orange box.
            \textbf{d}. Validation section. The longitudinal phase space is measured at screen~2 using S-band TCAV and spectrometer~\cite{diMitri2015layout} [\textbf{e}. measured longitudinal phase space for BC and XLIN OFF].
            Measured beam charge and reference energy for the reconstruction process are 250~pC and 260.5~MeV, respectively.
            }
   \label{fig:PALXFEL_Schematic}
\end{figure*}

A schematic of the GPSR technique is as follows [see Fig.~\ref{fig:PALXFEL_Schematic}\textbf{a}]: the initial 6-dimensional phase space distribution is parameterized by a generative machine learning model that is able to flexibly generate a wide variety of distribution structures.
The initially generated beam is then propagated through a simulated version of the beamline using a differentiable particle tracking software~\cite{gonzalez-aguilera:ipac2023-wepa065}.
To reconstruct the beam distribution, the parameters of the generative model are iteratively updated via gradient descent optimization~\cite{kingma2017adammethodstochasticoptimization} to minimize the difference between experimental measurements of the $(x,y)$ beam distribution on intercepting screens and corresponding simulated predictions.
Minimizing the difference ultimately yields a reconstructed beam phase space distribution that reproduces experimental measurements.

Previous experimental demonstrations of the GPSR method~\cite{PhysRevLett.130.145001,PhysRevAccelBeams.27.074601,PhysRevAccelBeams.27.094601} have shown accurate reconstruction of the phase space distribution validated by prediction of training and test datasets and comparison of beam parameters measured by conventional diagnostics.
Nonetheless, its broader applications at the accelerator facilities remains limited, as it still relies on specialized diagnostic elements such as TCAVs to rotate and probe the beam distribution in 6-dimensional phase space.
In addition, prior studies have primarily focused on validating reconstructions at a specific diagnostic point, leaving predictive accuracy and uncertainty of the reconstruction at additional downstream locations unexplored.

In this work, we demonstrate the first implementation and validation of the GPSR methodology for complete 6-dimensional phase space reconstruction using standard beamline elements commonly available at accelerator facilities~\cite{HUANG2021100097, conde_research_awa, PhysRevAccelBeams.22.101301}, including accelerating cavities and spectrometer magnets.
Through both simulation and experiment at the user facility, Pohang Accelerator Laboratory X-ray Free Electron Laser (PAL-XFEL)~\cite{app7050479, kang2017hard, nam2021high}, we show that a combination of quadrupole scans, dispersive measurements, and the accelerating cavity phase scans provides sufficient information to reconstruct the detailed 6-dimensional phase space distribution of the beam including complex and nonlinear features.
Furthermore, we conduct the first experimental validation  of the reconstruction methodology by accurately predicting independent, downstream experimental measurements of the beam phase space, such as a longitudinal phase space (LPS) and transverse-energy correlations at the end of the beamline (located approximately 27~m from the reconstruction region) excluded from the training process.
These downstream observations insofar are the best evidence of GPSR predictive accuracy, revealing that the reconstructed phase space closely approximates the ground truth physical distribution.
This work shows that the reconstruction framework enables efficient, high-fidelity 6-dimensional phase space diagnostics using only standard elements, and establishes a foundation for predictive diagnostics in a wide range of accelerator facilities.

\section{Results}
\subsection{GPSR methodology with standard beamline elements}
Previous 6-dimensional phase space reconstructions relied on the spectrometer magnets and TCAVs to directly measure the longitudinal correlations.
In contrast, we show in this demonstration that these correlations can be extracted using only the spectrometer and accelerating cavity readily available in most accelerator facilities.
Here, we should note that the methodology is not limited to this particular setup, but can be adapted to other standard beamline configurations that establish phase space correlations.

The key insight of the methodology can be conceptualized as a decomposition of the full reconstruction process into several components, each informed by physical measurements along the beamline.
First, the 4-dimensional transverse phase space $(x, x', y, y')_i$ is reconstructed using quadrupole scans and transverse beam distributions measured at a screen along the straight beamline~\cite{PhysRevLett.130.145001}.
Here, we note that 6-dimensional phase space of the beam is described by $(x, x', y, y', z, \delta)$ with $x'=p_{x}/p_{z}$, $y'=p_{y}/p_{z}$ where $p_{x,y,z}$ indicates normalized momentum in $(x,y,z)$ plane; $\gamma m_{e}\beta_{x,y,z}c$ normalized by $m_{e}c^{2}$ where $m_{e}$ indicates the electron mass.
$\delta=\left(p_{z} - p_{z,ref}\right)/p_{z,ref}$ where $p_{z,ref}$ is the reference particle's longitudinal momentum.
Subscripts $i$ and $f$ denote initial and final states, associated with the beam before and after the propagation to the beamline.
The reconstructed transverse phase space is then predominated and treated as constraints for dispersive measurements.
Accordingly, the horizontal beam distribution $x_f$ observed at the same screen with the spectrometer activated allows inference of the fractional energy spread $\delta_i$ through the dispersive measurement, as described in Eq.~\eqref{eq:dispersive_measurement}.
\begin{equation}
x_{f}=R_{11}x_{i} + R_{12}x'_{i} + R_{16}\delta_{i},
\label{eq:dispersive_measurement}
\end{equation}
where $R_{11}$, $R_{12}$ are transfer matrix components and $R_{16}$ is a dispersion, which are all determined by the beamline optics.
Therefore, through the GPSR process with measurements in straight line and dispersive section, 5-dimensional phase space $(x, x', y, y', \delta)_{i}$ becomes resolved.

To recover the longitudinal coordinate $z_i$, we utilized the energy modulation generated by the accelerating cavity. 
As the beam propagates through the cavity, its energy evolves as
\begin{equation}
E_{f}=E_{i} + eV\sin{(k_{RF}z_{i} + \phi_{RF})},
\label{eq:energy_cavity}
\end{equation}
where $k_{RF}$ and $\phi_{RF}$ are wavenumber and phase of the accelerating cavity, respectively.
$E_{i}$ and $E_{f}$ are initial and final energy distributions.
By activating or deactivating the accelerating cavity, we can clearly obtain the change of the beam energy $\Delta E$ that is related to the fractional energy spread $\delta_{i}$.
This enables reconstruction of the 1-dimensional projected $z_{i}$ distribution associated with $\delta_{i}$.

To further resolve the correlation between $z_{i}$ and $\delta_{i}$ including longitudinal phase space information, we perform RF phase scans.
These phase scans modulate the energy gain along the beam as depicted in Fig.~\ref{fig:PALXFEL_Schematic}\textbf{b}~\cite{1288795, DOWELL2003331}, leading to the longitudinal phase space rotation that allows the extraction of second-order moments~\cite{PhysRevSTAB.6.034202}.
This procedure is conceptually analogous to quadrupole scans in the transverse plane, where beam envelope evolution is used to calculate the transverse emittance.
Finally, by incorporating the quadrupole and cavity phase scans into the GPSR training process, we can achieve full 6-dimensional phase space reconstruction using only standard beamline elements.

In particular, since the 5-dimensional phase space is explicitly determined by measurements, the reconstruction of the longitudinal coordinate $z_{i}$ is strongly constrained by the intrinsic $(z, \delta)$ correlations described in Eq.~\eqref{eq:energy_cavity}.
Thus, this reconstruction process enables recovery of additional correlations associated with $z_{i}$, yielding a nearly unique solution that closely approximates the 6-dimensional phase space of the beam.
Thus, it has direct implications regarding prediction accuracy and uncertainty of the reconstruction.
We will investigate these factors with independent downstream measurements in the following sections.

\subsection{Experimental setup}
A schematic of the beamline elements used in the demonstration is shown in Fig.~\ref{fig:PALXFEL_Schematic}\textbf{c}.
We used an X-band RF accelerating cavity (XLIN) to obtain controlled energy modulations~\cite{PhysRevAccelBeams.28.012802}, which is typically used to linearize the non-linear energy chirp~\cite{emma2005xband} in XFEL facilities.
For the transverse phase space rotation, we used Q1 quadrupole magnet.
We used the first two dipole magnets in the bunch compressor (BC) section as the spectrometer.
Particularly, the BC is mounted on the movable stage support, allowing it to be configured into a straight line~\cite{Lee:IPAC2015-WEPMN041}. 
In this configuration, we can measure the projected $(x,y)$ beam distributions at the straight and dispersive sections with the screen~1 (See Methods for more details on the experimental setup). 

Downstream of the GPSR section comprises 3~quadrupole magnets, S-band TCAV, spectrometer magnet, and the screen.
This section is mainly used to measure the longitudinal phase space and other position-momentum correlations to investigate the prediction accuracy with independent downstream measurements.
It should be highlighted that that all downstream measurements are not included in the training process and used solely for validation.
Total length of the beamline from the Q1 to screen~2  is roughly 27~m.

\begin{figure*}[ht]
   \centering
   \includegraphics[width= 17.5 cm]{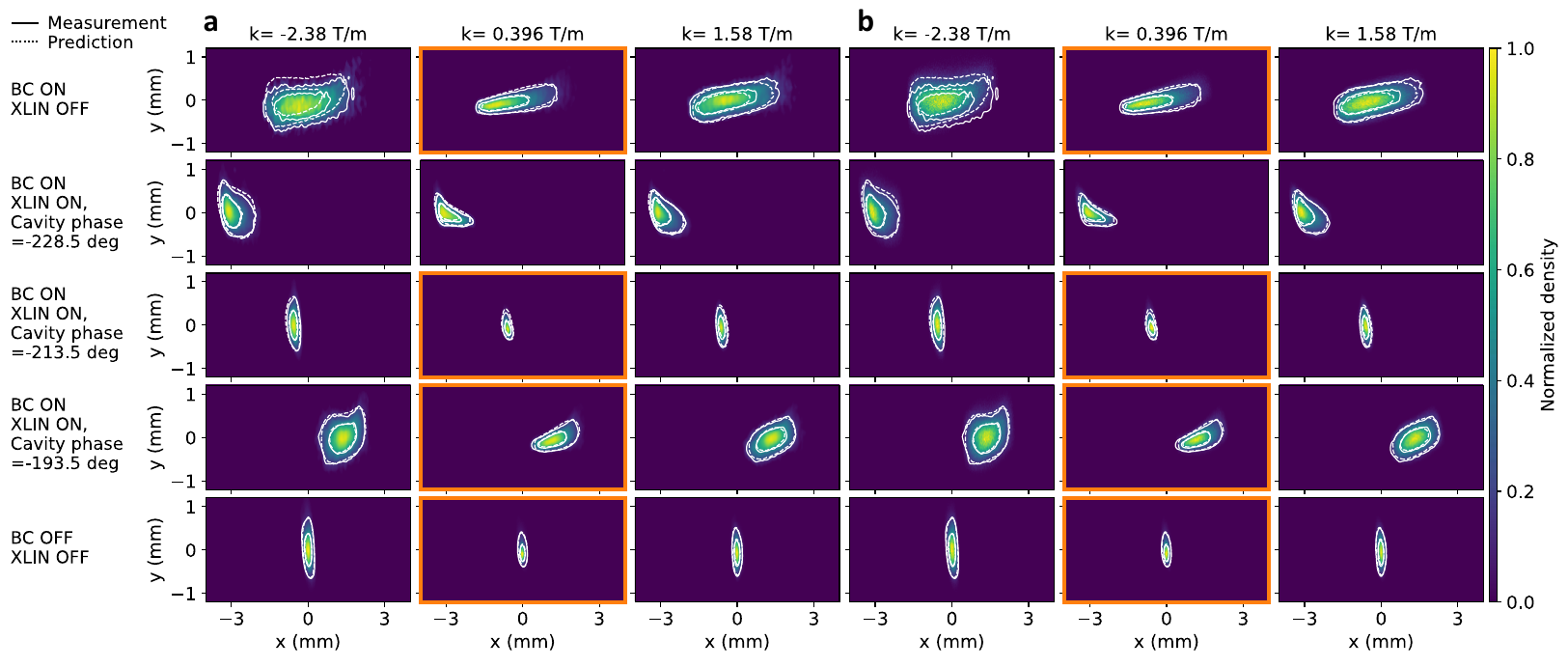}
   \caption{Comparison between measured beam distributions and predictions at the screen~1. Training dataset is highlighted with orange box.
   $k$ indicates the quadrupole magnetic field strength in T/m.
   Contour lines indicate 90 and 50th percentiles, where solid(dashed) lines represent the measurement(prediction).
   Background image shows the measurement for \textbf{a}, while it illustrates the prediction for \textbf{b}. }
   \label{fig:TestDataComparison_EXP}
\end{figure*}

In the data acquisition, we sampled beam distributions using a combination of 16 Q1 magnet strengths and 8 XLIN phase settings (see Methods for more details of the setup).
An additional setting with the XLIN cavity turned off was included to obtain the initial $(z,\delta)$ correlations of the beam before manipulation by the XLIN cavity.
Accordingly, there are a total of 9 XLIN phases.
For each beamline configuration, we took 15 shots at a repetition rate of 2~Hz with a charge window within $\pm3$\% of the nominal bunch charge of 250~pC.
The measured images were then averaged, and then resulting beam distribution was used as a data sample for the reconstruction process.
The total number of data samples was $16\times9\times2=288$, where the factor of 2 corresponds to the spectrometer configurations (BC ON/OFF).
It was then split in half to create a training dataset (144 samples) which was used to reconstruct the beam distribution, and a test dataset used to validate the accuracy of reconstruction predictions.

The training dataset generated during the experiment was then combined with a differentiable model. 
The differentiable beam dynamics simulation was built using the {\sc Bmad-X}~\cite{gonzalez-aguilera:ipac2023-wepa065} Python library, which implements up to second order beam tracking through the RF and magnetic elements using backwards differentiable calculations in PyTorch \cite{NEURIPS2019_bdbca288}. 
In this simulation, collective effects such as space charge or coherent synchrotron radiation (CSR) effects were neglected, as simulations using the {\sc elegant} code~\cite{flexible_sdds_code_accel_sim} indicated that these effects have negligible impact in this region of the beamline.

Initial training was performed using the full set of 144 training samples with 2,400 iterations (See Methods for details of the reconstruction setup).
We subsequently confirmed that the comparable GPSR performance can also be achieved using a reduced dataset and training iterations; it took approximately 30~minutes for 1,000~iterations on an NVIDIA A100 GPU at NERSC~\cite{NERSC2022}, where we used the dataset from only 5 XLIN phases (80 samples).
Accordingly, we will show the reconstruction results using reduced number of data samples.

\begin{figure*}[ht]
   \centering
   \includegraphics[width= 17.5 cm]{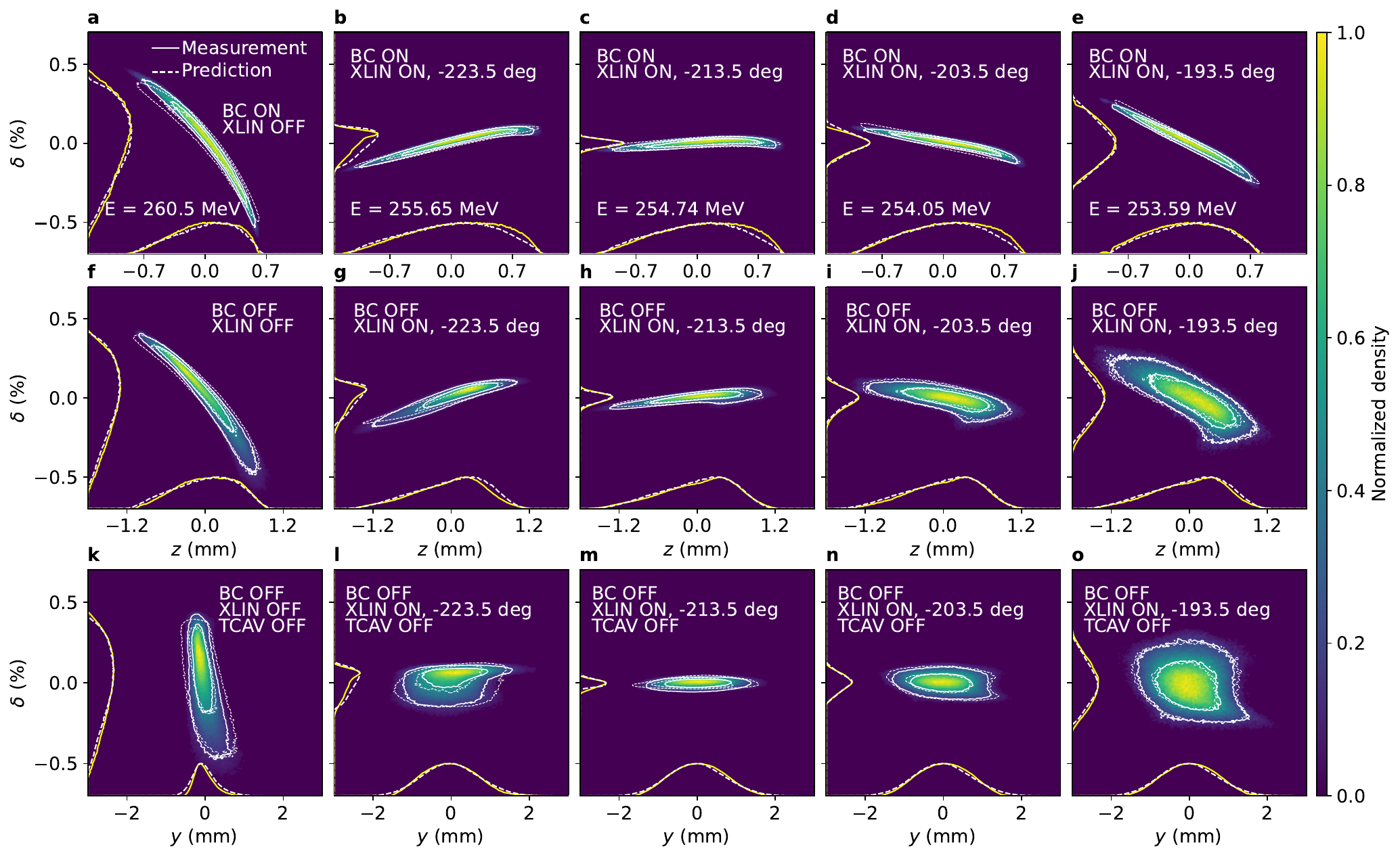}
   \caption{Comparison of LPSs and $(y,\delta)$ correlations at the screen~2. 
   In case of the LPS, the beam head is placed at positive $z$ while the tail is at negative $z$.
   Background image indicates the measured distribution.
   The average kinetic energy is indicated in each sub-figure, as the $\delta$ is distributed with respect to zero.
   White dashed(yellow solid), projected histogram indicates predicted(measured) density distribution.
   White contour lines indicate 90 and 50th percentiles. 
   Top row \textbf{a}-\textbf{e}: LPS where BC ON.
   Middle row \textbf{f}-\textbf{j}: LPS where BC OFF.
   Bottom row \textbf{k}-\textbf{o}: $(y,\delta)$ correlations with BC OFF.}
   \label{fig:LPS_YPZCOR_Comparison}
\end{figure*}

\subsection{Prediction of training and test datasets from the reconstructed phase space}
Prior to the experimental demonstration, we conducted a simulation-based validation of the reconstruction methodology using the beamline configuration described above.
We evaluated the accuracy on three representative 6-dimensional phase space distributions: i) a coupled Gaussian beam distribution, ii) a nonlinearly structured beam containing a double-horn energy profile, and iii) a realistic beam distribution obtained from upstream beam dynamics simulations of the nominal PAL-XFEL beamline. 
We found that the reconstruction methodology under the similar parameter scan conditions as in the experiment was able to accurately reconstruct qualitative and quantitative features of the beam.
The simulated data analysis are shown in the Supplementary Figs.~4-15.
In addition, we emphasize that the prediction of particle samples from a 6-dimensional phase space distribution is produced through the reconstruction process, which can be used to make predictions of beam dynamics or beam properties at different beamline settings or at different locations along the beamline.

First, Fig.~\ref{fig:TestDataComparison_EXP} shows a comparison between experimental measurements and reconstruction predictions for a subset of training and test settings at the screen~1.
In this figure, the background image in \textbf{a}(\textbf{b}) represents the measurement(prediction) [see Supplementary Figs.~1-2 for comparisons of full datasets].
In case of the dispersive measurements with the XLIN turned off (top row), small discrepancies are observed between the predicted and measured distributions, likely due to the low signal-to-noise ratios (SNR) or RF phase drift of accelerating cavities upstream of the demonstration section.
However, these small deviations do not compromise the ability of the trained model to capture detailed features of the beam, and they further demonstrate the robustness of the methodology in the presence of measurement noise.

Overall, the reconstructed beam distribution successfully reproduces shape and density of the measured distributions, including nonlinear features such as asymmetric tails.
In addition, we observed a clear modulation of the beam centroid position as the XLIN cavity phase varies.
This modulation indicates the energy-dependent change in beam trajectory with the fixed magnetic field of the dipole magnets, which highlights capabilities of the reconstruction methodology to infer the energy changes associated with the longitudinal coordinate $z_{i}$ from transverse beam measurements only.

\subsection{Validation of the methodology via independent downstream measurements}
In addition to analyzing the training and test datasets, a critical component of validating the methodology is the comparison of reconstructed beam characteristics with independent downstream measurements that are completely excluded from the training process.
We propagated the reconstructed 6-dimensional distribution through the remainder of the beamline using the {\sc elegant} simulation code, which takes into account collective and nonlinear optical effects such as coherent synchrotron radiation and chromatic aberrations.
The resulting predicted beam distributions were then compared with experimental measurements at the screen~2, located 27~m downstream from the reconstruction point. 
These measurements were obtained using a combination of quadrupole magnets, S-band TCAV, and spectrometer dipole as shown in Fig.~\ref{fig:PALXFEL_Schematic}\textbf{c}.

Figure~\ref{fig:LPS_YPZCOR_Comparison} presents a comparison between simulated predictions and experimental measurements of the longitudinal phase space and $(y,\delta)$ correlations under a variety of beamline conditions (see Methods for details on screen~2 measurements and Supplementary Fig.~3 for additional comparisons).
The image observed at the screen~2 represents the projected $(z,\delta)$ distribution in $(x,y)$ plane by means of the TCAV and spectrometer magnet.  
This implies that focusing effects by quadrupole magnets located around the diagnostic region introduce additional transverse correlations on the beam, distorting the projected $(z,\delta)$ mapping.
Thus, the measured image is a result of both longitudinal and transverse phase space projections.

Despite these inherent complexities of downstream measurements, the reconstructed beam distribution accurately predicts the 1-dimensional projected histograms along the longitudinal coordinates $z$ and $\delta$, showing excellent agreement with experimental measurements.  
In addition, the 2-dimensional phase space densities obtained from the reconstruction almost match the measured distributions, successfully capturing nonlinear correlations and rotation of the longitudinal phase space.
The reconstruction also recovers key features of the beam before the manipulation by the XLIN cavity including bunch length compressions, when the entire bunch compressor section is active (see horizontal axis $z$ in Fig.~\ref{fig:LPS_YPZCOR_Comparison}\textbf{a-j}).

Furthermore, the reconstruction successfully reproduces transverse-longitudinal correlations $(y,\delta)$ as shown in Fig.~\ref{fig:LPS_YPZCOR_Comparison}\textbf{k–o} [see also Supplementary Figs.~6, 10, and 14, which illustrate other correlations from simulated demonstrations, such as $(x,\delta)$ and $(x',z)$].
Although slight discrepancies between predicted and measured 1- and 2-dimensional projections are observed in some cases, these are minor relative to the overall beam structure.  
Whether such deviations originate from the trained model or imperfections in the simulation of the beamline compared to the actual experimental setup remains a subject of future works.

Nevertheless, these results confirm that the reconstruction methodology is able to accurately predict independent downstream measurements, demonstrating not only the robustness of the predictive diagnostics, but also that the initially reconstructed distribution can be considered close approximation to the actual physical beam.
Thus, this validation indicates that predictive diagnostics are applicable at any beamline locations along the reconstruction section. 
In addition, the reconstructed phase space information can be used either as a target or as an input to calibrate the simulated model of the upstream beamline and further downstream lattice for the predictive diagnostics.

\begin{figure*}[ht]
   \centering
   \includegraphics[width=17.5 cm]{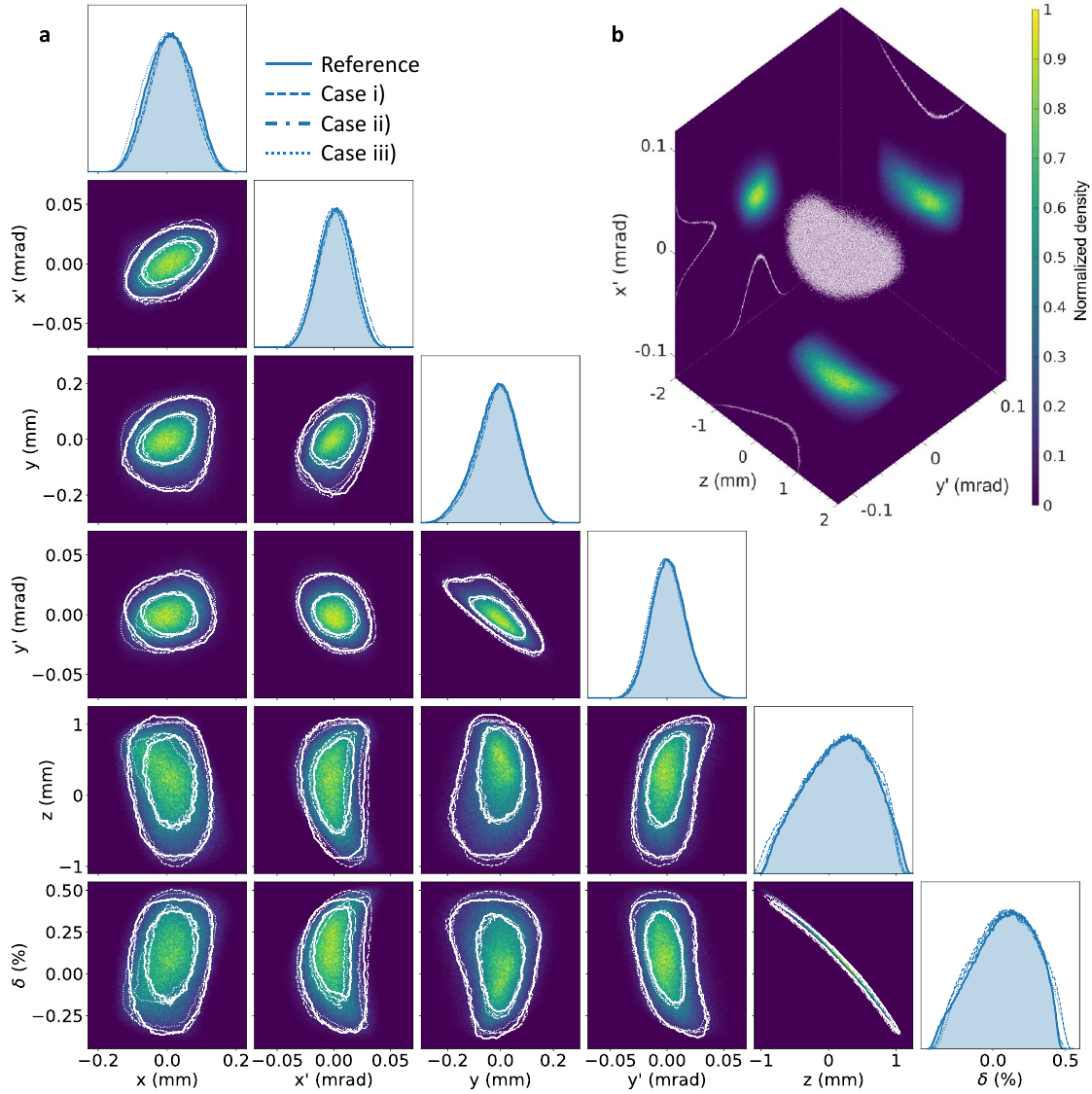}
   \caption{\textbf{a}. Reconstructed 6-dimensional phase space. Contour lines indicate 90 and 50th percentile, while different line styles represent three cases in the error study. "Reference" indicates the nominal case. Note that only 90th percentile is shown for $(z-\delta)$ distribution. 
            Blue histogram indicates the 1-dimensional histogram in each coordinate.
            \textbf{b}. 3-dimensional scatter/density plot of the reconstructed beam phase space.
            }
   \label{fig:6D_EXP_Reconstruction}
\end{figure*}

\subsection{Reconstructed 6-dimensional beam phase space}\label{6dphase}
Finally, Fig.~\ref{fig:6D_EXP_Reconstruction} illustrates the reconstructed 6-dimensional phase space of the beam at the entrance to the Q1 magnet.
Figure.~\ref{fig:6D_EXP_Reconstruction}\textbf{a} shows 2-dimensional projections of the beam distribution, providing a comprehensive view of the correlations among all beam coordinate axes, while Fig.~\ref{fig:6D_EXP_Reconstruction}\textbf{b} shows a subset of 3-dimensional projections along coordinate axes.
The reconstruction predicts non-linear tails in the transverse dimensions and  accurately reproduces the nonlinear energy chirp, as observed by direct measurements shown in Fig.~\ref{fig:LPS_YPZCOR_Comparison}.
In case of the phase space regarding $\delta$, the center position is slightly shifted by 0.07\%, which corresponds to 0.2~MeV energy deviation compared to the reference energy of 260.5~MeV (See Methods on how to determine the reference energy).
Therefore, This reconstruction suggests that more accurate reconstruction could be achieved by applying a slight adjustment to the reference energy of the beam.

\begin{table}[bp]
\caption{
Comparison of the Twiss parameters, emittances, and RMS quantities.
Values in parenthesis show the averaged values with standard deviation $1\sigma$ for several different GPSR setup. See Sec.~\ref{6dphase} for more details on the setup.
}
\begin{ruledtabular}
\begin{tabular}{l r r r}
\textrm{Case} & \textrm{4D reconstruction} & \textrm{6D reconstruction} & \textrm{unit}\\
\colrule
$\epsilon_{nx}$ & 0.42 & 0.40 (0.42$\pm$0.01) & mm~mrad\\
$\epsilon_{ny}$ & 0.39 & 0.40 (0.39$\pm$0.01) & mm~mrad\\
$\beta_{x}$ & 4.53 & 4.67 (4.59$\pm$0.15)& m\\
$\beta_{y}$ & 7.78 & 8.12 (7.87$\pm$0.29) & m\\
$\alpha_{x}$ & $-0.56$ & $-0.50$ ($-0.52\pm$0.04) & rad\\
$\alpha_{y}$ & 1.25 & 1.17 (1.16$\pm$0.09) & rad\\
\colrule
\textrm{Case} & \textrm{Measurement} & \textrm{Reconstruction} & \textrm{unit}\\
$\epsilon_{nz}$ & 137.34$\pm$17.24 & 140.05 (141.29$\pm$6.27) & mm~mrad\\
$\sigma_{z}$ & $0.45\pm0.01$ & 0.49 (0.47$\pm$0.01) & mm\\
$\sigma_{\delta}$ & $0.24+0.01$ & 0.22  (0.22$\pm$0.00)& \%\\
\end{tabular}
\end{ruledtabular}
\label{table:beamparameters_comp} 
\end{table}

We also calculated ensemble-averaged quantities of the beam as summarized in Table~\ref{table:beamparameters_comp}.
These predicted values of the 6-dimensional reconstruction are quantitatively consistent within 10\% of those obtained by 4-dimensional phase space reconstructions using the GPSR analysis procedure outlined in Ref.\cite{PhysRevLett.130.145001}.
For the longitudinal phase space at the end of the beamline, the reconstructed values agrees well with conventional calculations of longitudinal bunch properties, such as the RMS longitudinal bunch length, momentum deviation, and emittance (where uncertainty estimates come from systematic uncertainties in image thresholding, see details in Methods).
Finally, the RMS slice energy spread$-$a critical quantity of interest for XFELs$-$is predicted to be 0.01\%, which is consistent with typical values observed during normal operations at the PAL-XFEL.

In addition, we investigated the uncertainty of the reconstructed phase space by performing multiple training runs under different configurations of the number of iterations and data samples.
We considered 3 cases additionally: i) training with all 144 datasets for 2,400 iterations, ii) training with the same datasets for 1,200 iterations, and iii) training with 80 datasets using different XLIN phase settings for 1,000 iterations.
The reconstructed beam parameters listed in Table~\ref{table:beamparameters_comp} in parentheses indicates the averaged values with standard deviation $1\sigma$.
In addition, contour lines in Fig.~\ref{fig:6D_EXP_Reconstruction}\textbf{a} show specific percentiles of the phase space distributions.
In the third case, correlations regarding $x$ are somewhat deviated compared to the other cases, implying that the chosen cavity phases may not have been optimal to get beam correlations during the training.
Nevertheless, the overall results show that the reconstruction methodology provides well converged values, exhibiting low sensitivity to random seeding variations.
Furthermore, we found that the predicted beam distributions for those cases also remain in qualitative agreement with the experimental measurements.

\section{Discussions}
We have validated prediction accuracy of the GPSR methodology using standard accelerator elements, including the accelerating cavity and spectrometer magnet, without relying on specialized deflecting structures to extract $(z,\delta)$ correlations.
The reconstructed 6-dimensional phase space not only shows excellent agreement with the training and test datasets, but also remains in quantitatively and qualitatively agreement with measured beam distributions further downstream in the beamline, which are excluded from the reconstruction training.
These observations provides strong evidence that the GPSR can be effectively applied to systems that establish transverse–longitudinal correlations, regardless of the specific source, thereby greatly enhancing its flexibility and accessibility for beam diagnostics.

The reconstruction methodology offers broad applicability across accelerator science. 
In particular, the information of high-fidelity beam phase space can be used with non-differentiable simulations that consider collective and nonlinear effects. 
This compatibility extends its utility to a wide range of challenges in accelerator development, including beamline optimization, predictive beam characterization along multiple beamline location with forward-backward particle trackings, and model calibration for future digital twin frameworks~\cite{Edelen:2024iuj}.
In the context of FEL facilities, the reconstructed phase space can be applied to the measurement-based studies on suppression of emittance growth induced by the CSR effect and investigation of matching conditions along the beamline, thus to eventually enhance the FEL quality.
Moreover, in advanced accelerator concepts where beam quality preservation is one of the most important tasks~\cite{PhysRevLett.126.014801,Lindstrom2024}, the optimization of conditions for injection into wakefields (either based on structures or plasmas) becomes critically important. 
The present framework provides a powerful diagnostic tool to support such optimization, thereby enhancing the performance of the future advanced acceleration schemes.

In order to apply the GPSR method for the low-energy beam, consideration of space charge effect is necessary.
Recently, implementation of the space charge kick matrix into the differentiable simulation is under development~\cite{cosgun2025ipac}.
We will further investigate reconstruction for space charge-dominated beams.
In addition, in the case of the uniqueness of the beam phase space through the reconstruction process, we will perform additional dedicated investigations to enhance the accuracy and reliability of the reconstruction methodology.

The GPSR methodology presented in this work paves the way for significant advances in the realization of predictive beam diagnostics using readily available elements in a variety of accelerator facilities such as accelerator-based light sources, centers for the advanced accelerator concepts, and beam test facilities.

\section{Methods}
\subsection{PAL-XFEL}
The PAL-XFEL facility consists of a photoinjector, 3~bunch compressor sections, and number of accelerator columns.
The photoinjector comprises 2.856~GHz, 1.6~cell copper-cathode RF gun and two 3.0~m accelerating cavities~\cite{Hong_Han_Park_Jung_Kim_Kang_Pflueger_2015}.
Wavelength of the UV laser illuminating the cathode is 257~nm.
UV laser distribution is basically Gaussian shape in both transverse and longitudinal planes, but truncated in the transverse plane to obtain low emittance against the space charge force along the injector section.
A full width at half maximum UV length was 2.8~ps.
Nominal beam charge is set to 250~pC during normal operation for user service.
When the beam is generated along the photoinjector, it is propagating through the laser heater and Hard X-ray Linac~1 (HL1) section as shown in Fig.~1 of Ref.~\cite{kang2017hard}.
The main purpose of the laser heater is to increase the slice energy spread of the beam and suppress a micro-bunching instability. 
The HL1 section consists of four 3.0~m accelerating cavities, with two of them interconnected as single module.
In this demonstration, we turned off the second accelerating module to only use the XLIN as the accelerating cavity.
The kinetic energy of the beam becomes reduced to 260.5~MeV.

\subsection{Setup for data acquisition and GPSR training}
In the demonstration section, operational phase and length of the XLIN cavity is 11.424~GHz and 0.6~m.
The bending angle and arc length of the bunch compressor dipole magnet are $4.97$~deg and 0.2~m, respectively.
The distance between Q1 magnet and screen~1 is about 12.6~m.
In case of the downstream validation section, the operational frequency of 1.066~m S-band TCAV is 2.856~GHz, where the spectrometer dipole magnet has the bending angle and arc length of 15~deg and 0.7~m.

We set the range of the Q1 magnet strength from $-2.77$ to $3.16$~T/m with 16 samples to get the nominal parabolic beam envelope used for the conventional quadrupole scan. 
We also set the XLIN cavity phase from $-228.5$ to $-193.5$ deg with 5-degree steps including XLIN turned off.
In this range of the cavity phase, changes of the energy chirp were clearly captured as reported in Ref.~\cite{DOWELL2003331}.
The initial beam energy before modulation by the XLIN was measured at the screen 2 using the spectrometer magnet.
Likewise, the energy change due to the XLIN cavity was measured at the same screen.
While we referred to these values as the initial condition of the GPSR process, we found that the reconstruction suggests the reference beam energy and XLIN voltage to be 260.5~MeV and 8.4~MV/m, respectively.
In this setup, the reconstructed phase space is obtained where the center of each phase space is placed close to zero (e.g., $\braket{\delta}\sim0$).

In case of the dipole magnet modeling, a linear fringe field integral (FINT) was used to consider the fringe field~\cite{brown1975slac}. 
Whereas, a hard-edge field in the quadrupole magnet was adapted since the fringe field effect is negligible in our setup.
During the data acquisition (DAQ), the Q2 magnet was fixed at a strength of $+2.54$~T/m while Q$3-4$ were turned off.
This setting ensured that the beam size at the center of the Q1 scan range was minimized in both horizontal and vertical planes, thus forming quadratic response of the beam envelope.
In addition, under these conditions, we measured the fractional energy spread distribution while minimizing the horizontal beam size contribution~\cite{PhysRevAccelBeams.21.062801}.

In the GPSR process, the initial phase space distribution was sampled from a multivariate normal distribution where the scaling factor was set to 0.001 to ensure that the initial 6-dimensional phase space is on the order of mm in position and mrad in momentum. 
We used fully connected neural networks with two hidden layers connected by Tanh activation functions, where each layer contains 20 neurons. 
The back-propagation, based on the gradient-descent method where the learning rate was set to 0.01, was performed using a mean absolute error-based loss function estimated by the comparison between experimental measurements and predictions~\cite{PhysRevAccelBeams.27.094601}.

\subsection{Beam measurements in dispersive section}
During measurements in dispersive section for the reconstruction model training, a current, corresponding to the magnetic field of the BC magnets, was fixed.
The change of the beam energy induced by the XLIN cavity with respect to the RF phase results in not only the evolution of the beam shape but also the shift of the beam centroid position at the screen~1.
Accordingly, we set the procedure for the experimental data acquisition as follows: i) we first found the phase of the XLIN that provides the minimum horizontal beam size at the screen~1. 
This is the case where the energy chirp is minimized. ii) We then set the BC dipole current to adjust the beam position to be at the center of the screen~1; this is a procedure of defining the reference case when the XLIN is turned on.
At the same time, iii) we fixed the BC dipole magnetic field for data acquisition.
Finally, iv) we set the XLIN phase range that provides the beam images that are positioned within the given aspect ratio of the screen~1.

\subsection{Image processing and downstream measurements}
When the BC is turned on, the projected beam distribution at the screen~1 spreads due to the energy correlation.
Thus, with given amount of scintillating light, background noise becomes relatively large.
To mitigate the noise, we averaged all 15 images and subtracted the background noise.
The signal-to-noise ratio (SNR) value, defined as a ratio between maximum and background intensity of the image, varies depending on the beam size.
The lowest observed SNR was approximately 1.6 for the case with the BC ON and XLIN OFF, which corresponds to the largest beam distribution in the dataset. 
Except for this case, the SNR value was generally larger than 3.0.
In addition to removing the noise, we applied a Gaussian filter with the standard deviation of 2 pixels and a median filter with the kernel size of 2 to smooth the beam distribution.
We also performed the threshold subtraction using the triangle threshold method~\cite{zack-1977-a}, with the multiplication factor of 1.7.
The image at the screen~1 was cropped to 470$\times$260 with the pixel resolution of roughly 20.2~\si{\micro\meter}/pixel due to limited memory (40~GB) of the GPU during the GPSR training.

In case of the screen~2, the calibrated energy resolution (in horizontal axis) with respect to the measured dispersion is approximately 0.004\%/pixel.
Likewise, the estimated TCAV calibration factors were roughly 4.8~\si{\micro\meter}/pixel for BC OFF and 4.4~\si{\micro\meter}/pixel for BC ON according to the TCAV voltage set to 1.7(1.1)~MV for BC ON(OFF).
These TCAV voltage values were used in the beam dynamics simulation with the reconstructed phase space to predict the LPS and other transverse-energy correlations which are measured downstream at the end of the beamline.
During these measurements, while Q$3-4$ were turned off, Q$5-7$ were activated with the field strengths of $-0.89$, $-2.43$, and $5.43$~T/m to properly propagate the beam to the screen~2.

Due to the low SNR of the screen~1 image, we suspected that some of the beam distributions were unintentionally cut off during the image processing. 
In contrast, the SNR of the screen~2 image was typically larger than 100.
As a result, the measured LPS and $(y,\delta)$ correlations without threshold subtraction led to underestimation of the predicted distributions.
Therefore, we applied finite level of the threshold value to the screen~2 image to attenuate the beam signal and compare it with the prediction. 
When the BC was turned off, the threshold level of the LPS and $(y,\delta)$ images were set to approximately $10-12.5$\% of the maximum intensity of the image.
Similarly, in the case when the BC was turned on, the threshold level of the LPS image was set to roughly 22.5\% of the maximum intensity.
The error of the longitudinal emittance was calculated using 3 cases with different threshold levels; $\pm$2.5\% based on the reference threshold value of 10\%.

\section{Data availability}
Software and example of simulated demonstration are available at \cite{gpsr_example}.
The experimental data presented in this paper can be made available from the corresponding authors upon reasonable request.

\section{Code availability}
This research used the {\sc Bmad-X} code~\cite{gonzalez-aguilera:ipac2023-wepa065} for differentiable simulations, which is available in the public repository https://github.com/bmad-sim/Bmad-X. 
The GPSR framework for the phase space reconstruction is available in the github repository https://github.com/roussel-ryan/gpsr.

\section{Acknowledgements}
Authors thank PAL-XFEL accelerator control team members and Prof. Moses Chung for support on the accelerator control and and fruitful discussions on beam correlations in early stage of the study. This work was supported by the National Research Foundation of Korea (NRF) grant (2019R1I1A1A01041573, RS-2024-00347026, RS-2024-00455499), funded by the Korea government (MSIT). This work was also supported by the U.S. Department of Energy, Office of Science under Contract No. DE-AC02-76SF00515 and the Center for Bright Beams, NSF award PHY-1549132.This research used resources of the National Energy Research Scientific Computing Center (NERSC), a U.S. Department of Energy Office of Science User Facility located at Lawrence Berkeley National Laboratory, operated under Contract No. DE-AC02-05CH11231 using NERSC award BES-ERCAP0020725.

\section{Author contributions}
S.K. and R.R. led the conception of the GPSR methodology with standard accelerator elements.
J.P.G.A. and R.R. developed the differentiable simulations and GPSR framework and Y.K.K. provided guidance and oversight.
S.K. led the conceptual study including simulated demonstrations.
S.K., G.K., and H.Y. performed the experiment with support from M.-H.C., C.H.S., and H.H.
S.K. led the data analysis including GPSR training, tracking simulation, and data visualization.
J.P.G.A., R.R., G.K., A.E., and H.Y. contributed to the data analysis and validation.
S.K. and R.R. wrote the draft manuscript and S.K., R.R., and A.E. edited the manuscript.
All authors reviewed and agreed to the final version of the manuscript.

\section{Competing interests}
The authors declare no competing interests.


%

\end{document}